\documentclass[journal=nalefd,manuscript=article,amsmath,amssymb]{achemso}

\usepackage[T1]{fontenc} 
\usepackage{graphicx}
\usepackage{amssymb}
\usepackage{amsfonts}
\usepackage{amsmath}
\usepackage{amsbsy}
\usepackage{natbib}
\usepackage{comment}
\usepackage{color}

\author{Xi-guang Wang}
\affiliation[csu]
{ School of Physics and Electronics, Central South University, Changsha 410083, China}
\alsoaffiliation[martinluther]
{Institut f\"ur Physik, Martin-Luther Universit\"at Halle-Wittenberg, D-06120 Halle/Saale, Germany}
\author{Guang-hua Guo}
\affiliation[csu]
{ School of Physics and Electronics, Central South University, Changsha 410083, China}
\author{Jamal Berakdar}
\email{jamal.berakdar@physik.uni-halle.de}
\affiliation[martinluther]
{Institut f\"ur Physik, Martin-Luther Universit\"at Halle-Wittenberg, D-06120 Halle/Saale, Germany}

\title {Steering  magnonic dynamics and  permeability at exceptional points  in a  parity-time symmetric  waveguide}
\abbreviations{IR,NMR,UV}
\begin{document}
%
%
%
\begin{abstract}
Tuning the low-energy magnetic  dynamics  is a  key element in designing  novel magnetic metamaterials, spintronic devices and  magnonic logic circuits.
This study uncovers  a new, highly effective way of controlling the magnetic permeability via shaping  the  magnonic properties in coupled magnetic  waveguides separated by current carrying spacer with strong spin-orbit coupling. The spin-orbit torques exerted on the   waveguides leads to an externally  tunable   enhancement of  magnetic damping in one waveguide and a decreased damping in the other, constituting so a magnetic  parity-time  (PT) symmetric system  with emergent  magnetic properties at the verge of the exceptional point where   magnetic gains/losses are balanced.  In addition to controlling the magnetic permeability,   phenomena inherent to PT-symmetric systems are identified,   including the  control on
  magnon power oscillations,  nonreciprocal magnon propagation, magnon trapping and enhancement as well as the increased sensitivity  to magnetic perturbation and abrupt spin reversal. These predictions are demonstrated  analytically and confirmed by 
  full numerical simulations under experimentally feasible conditions. 
 The  position of the exceptional points and the  strength of the spontaneous PT symmetry breaking can be tuned by external electric and/or magnetic fields. 
The roles  of the intrinsic magnetic damping, and the possibility of an electric control via  Dzyaloshinskii-Moriya interaction are exposed  and utilized for mode dispersion shaping and magnon amplification and trapping. 
The results point to a  new route to designing optomagnonic waveguides, traps, sensors, and circuits.
\end{abstract}


\section{Introduction}
\begin{figure}
	\includegraphics[width=0.6\textwidth]{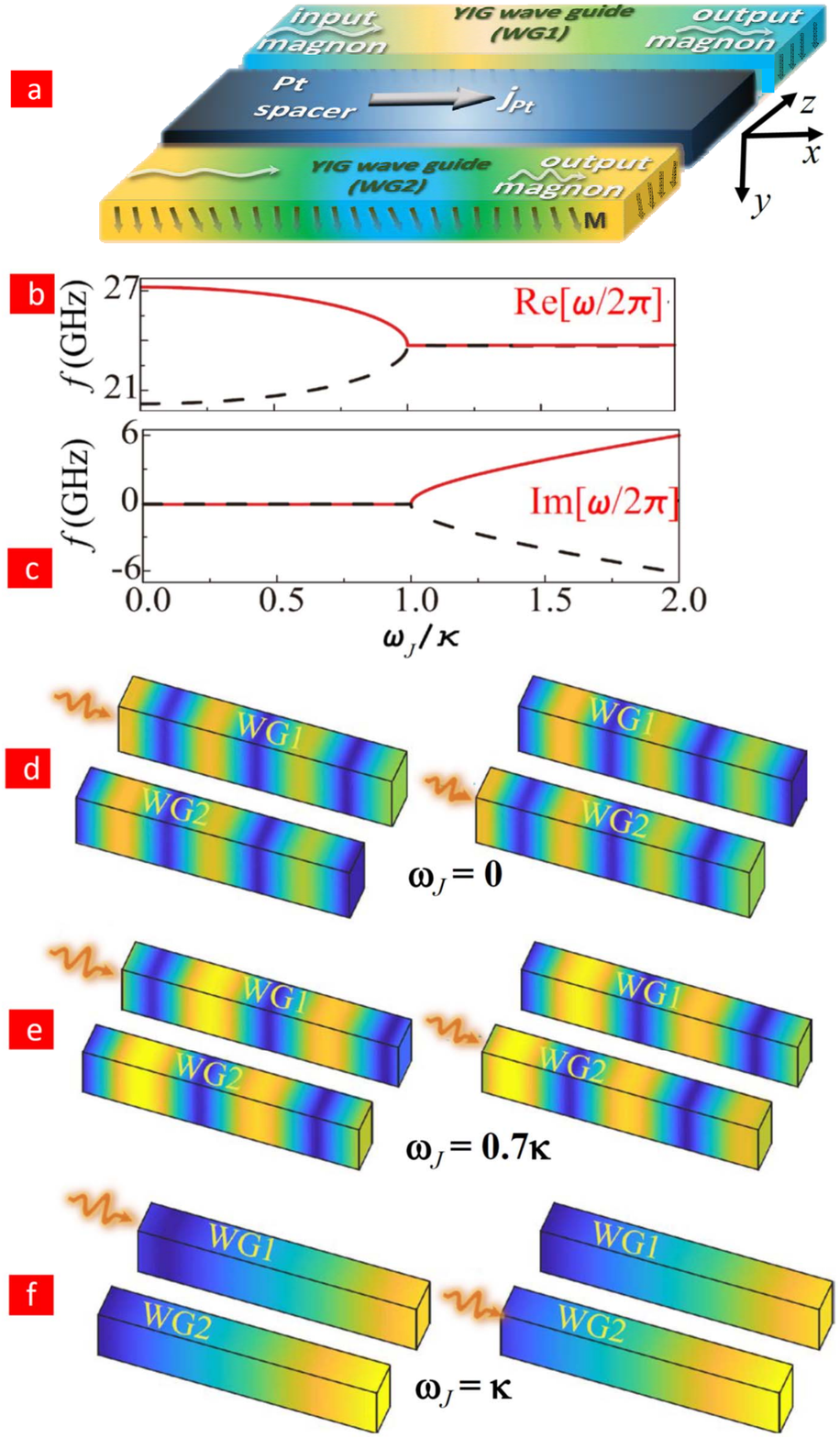}
	\caption{\label{model} (a)  Two magnetic waveguides   (labeled as WG1 and WG2, as an example we use  YIG in the numerical  simulations)
		are coupled via RKKY  interaction with metallic spacer that has a large spin Hall angle (here Pt). Driving a charge  current $ \vec{j}_{\mathrm{Pt}} $ along the space  ($x$ direction)
		results in spin-Hall torques acting on the magnetic waveguides. The torques damp or antidamp the magnetic dynamics in WG1 and WG2  resulting so in PT-symmetric structure with a  new (PT) symmetry behavior of the magnetic permeability. 
		Magnon wave packets are launched locally   at one end of WG1 or WG2 (left side in the figure)  and the propagation characteristics of the magnonic signal is steered, amplified or suppressed by external fields that drive the waveguides from the PT-symmetric to the PT-symmetry broken phase through the exceptional point where magnetic losses in WG1 balance magnetic antidamping in WG2. This can be achieved for instance by changing the ratio between the intrinsic coupling strength between the waveguides $\kappa$ and the strength of the  spin-Hall torques $\omega_J$.  (b) Real and (c) imaginary parts of two eigenmode frequency 
		 $f= \omega/(2\pi) $ as we scan   $ \omega_J/\kappa $ at the wave vector $ k_x = 0.1 $ nm$ ^{-1} $. (d-f) Spatial profiles of propagating spin wave amplitude for different loss/gain balance  (different $ \omega_J $), when the spin waves are locally excited either in WG1 or WG2. The color change from blue to red corresponds to a linear amplitude change ranging from 0 to the maximum of input signal. The local microwave field excites spin waves at the left side of the waveguide and  has a frequency of 20 GHz. The length (along $ x $ axis) of waveguides in (d-f) is 580 nm.}
\end{figure}

Nanomagnetism  is the backbone of spin-based  memories, data processing and sensorics.   In a generic magnet, the permeability, meaning the magnetic response to a weak  external perturbation is governed by the  behavior of the  spin waves which are  collective transverse oscillations (with their quantum termed magnon)  around the ground state. Miniaturized 
  magnonic logic circuits   \cite{Chumak:2015fa,Hillebrand2,Kruglyak,Vogt,Chumak,Sadovnikov:2015hm}  and waveguides  operated at low energy cost with negligible  Ohmic losses were demonstrated. Furthermore,  geometric confinements, nanostructuring,  and material design allow a precise  spectral shaping and guiding of magnons, which is reflected respectively in a modified magnetic response. 
  Here we point out   an approach based on a magnonic gain-loss mechanism in two waveguides with a normal-metal spacer. The two magnetic waveguides  are coupled via the Ruderman-Kittel-Kasuya-Yosida  (RKKY) interaction.
   Driving  charge current  in a  spacer layer with a strong spin-orbit coupling (cf. \ref{model}(a)),   spin orbit torques (SOTs) are exerted on the magnetizations of the waveguides. In effect, SOT adds  to the intrinsic magnetic damping, as evident  from Eq. (\ref{LLG}).  It is thus possible to achieve a case where SOT-induced magnetic losses in one waveguide are balanced by antidamping in the other waveguide.  This is a typical case of a PT-symmetric system as realized for instance in optical systems \cite{Ruschhaupt_2005,Feng972,PhysRevLett.100.103904,Ruter2010,Regensburger2012,PhysRevX.8.021066}. A hallmark of PT-symmetric systems is that, even if the underlying Hamiltonian  is non-Hermitian, the eigenvalues  may be real  \cite{PhysRevLett.80.5243,PhysRevLett.89.270401,Bender_2007}, and  turn complex when crossing the "exceptional point"  and  entering the PT-symmetry broken phase upon varying a parametric dependence in the Hamiltonian.   Examples were demonstrated in optics and photonics \cite{Feng972,PhysRevLett.100.103904,Ruter2010,Regensburger2012,PhysRevX.8.021066,Peng2014,PhysRevLett.106.213901,Nazari:11,Kartashov:09,PhysRevA.86.023807, PhysRevLett.103.093902},
optomechanics \cite{PhysRevLett.114.253601,Xu2016}, acoustics\cite{PhysRevX.4.031042,Fleury2015} and electronics \cite{PhysRevA.84.040101,Schindler_2012,Assawaworrarit2017,Chen2018}.
Also,  PT-symmetric   cavity magnon-polaritons were discussed involving  phonon dissipation or electromagnetic radiation as well as  parametric driving or SOT effects  \cite{PhysRevB.91.094416, PhysRevB.94.020408, PhysRevB.97.201411, Zhang2017, PhysRevB.99.214415, PhysRevLett.121.197201}. \\

Our main goal is the design and demonstration of  PT-symmetric magnonic waveguides which are controllable by  feasible external means that   serve as a knob to tune the system across the exceptional point. In addition to the documented advantages of magnons, this would  bring about  new  functionalities that can be integrated in optomagnonic, spintronic, and magnonic circuits. 

\section{Spin-torque driven PT-symmetric waveguides}
Our magnon signal propagates along  two  magnetic  waveguides (which define the $\vec x$ direction) coupled via RKKY exchange interaction (cf. Fig, \ref{model}(a)).   A charge  current flowing in a spacer with a large spin Hall angle (such as Pt) applies  a SOT $ \vec{T}_1 \parallel \vec{y} $  on first waveguide enhancing the effective damping, and a SOT $ \vec{T}_2 \parallel -\vec{y} $ on second waveguide weakening the effective damping.  The polarization directions of the spin Hall effect induced transverse spin currents $ \vec{T}_1 = \vec{z} \times \vec{j}_{\mathrm{Pt}} $ in WG1 and $ \vec{T}_2 = (-\vec{z}) \times \vec{j}_{\mathrm{Pt}} $ in WG2, are related to  the charge current density  $ \vec{j}_{\mathrm{Pt}} $.  
 In a generic ferromagnet and for the long wavelength spin excitations of interest here, to describe the magnetic dynamics  it is sufficient to adopt a classical continuous approach and solve for the equations of motion of the magnetization vector fields $\vec{M}_p(\vec {r}, t)$ ($ p = 1,2 $ enumerates the two waveguides), which amounts 
to propagating the Landau-Lifshitz-Gilbert (LLG) equation,\cite{Krivorotov228,PhysRevLett.109.096602,Garello2013,Hoffmann2013}
\begin{equation}
\displaystyle \frac{\partial \vec{M}_p}{\partial t} = - \gamma \vec{M}_p \times \vec{H}_{\mathrm{eff,}p}+  \frac{\vec{M}_p}{M_{s}} 
 \times\left[\alpha  \frac{\partial \vec{M}_p}{\partial t} - {\gamma c_J}  \vec{T}_p \times \vec{M}_p\right].
\label{LLG}
\end{equation}
The waveguides  are  located at  $ z = +z_0 $ and $ z = -z_0 $.  We are interested in small transversal excitations and hence it is useful to use the unit vector field $ \vec{m}_p =\vec{M}_p/ M_s $ where $ M_s $ is the saturation magnetization and $\gamma$ is the gyromagnetic ratio. $ \alpha $ is the conventional  Gilbert damping inherent to magnetic loses  in each of the waveguides. The effective field $ \vec{H}_{\mathrm{eff,}p} = \frac{2 A_{\mathrm{ex}}}{\mu_0 M_s} \nabla^2 \vec{m}_p + \frac{J_{\mathrm{RKKY}}}{2\mu_0 M_s t_p} \vec{m}_{p\prime} + H_0 \vec{y}  $ consists of the internal exchange field, the interlayer RKKY coupling field, and the external magnetic field applied along the $ y $ axis, where $ p, p\prime = 1,2 $, and
 $ p\prime \ne p $. $ A_{\mathrm{ex}} $ is the exchange constant, $ J_{\mathrm{RKKY}} $ is the interlayer RKKY exchange coupling strength, 
 $ t_p $ is the thickness of the $ p $th layer, and $ \mu_0 $ is the vacuum permeability.  Of key importance to this study is the strength $ c_J = T \theta_{\mathrm{SH}}\frac{ \hbar\; J_e}{2\mu_0 e \; t_p M_s} $ of SOT which is proportional to charge-current density $ J_e $ and the spin Hall angle $ \theta_{\mathrm{SH}} $ in the spacer layer, 
 for instance, at the exceptional point defined in the following study, $c_J = 1\times 10^5 $ A/m corresponds to a charge current density of $J_e = 9 \times 10^{8} $ A/cm$ ^2 $ in Pt \cite{Collet2016}.  $ T $ is the transparency at the interface, and $ e $ is the electron charge.
 Our proposal applies to a variety of settings, in particular synthetic antiferromagnets \cite{syAFM} offer a good range of tunability.
  To be specific, we present here  numerical simulations for  Pt interfaced with  a Yttrium-Iron-Garnet (YIG) waveguides as experimentally realized  for instance in Ref. [\citenum{Collet2016}] corresponding  to the following values $ M_s = 1.4 \times 10^5 $ A/m, $ A_{ex} = 3 \times 10^{-12} $ J/m (technical details of the numerical realization are in the supplementary materials). For the Gilbert damping we use $ \alpha = 0.004$ but note that  depending on the quality of the waveguides $\alpha$ can be two order of magnitude  smaller.  The interlayer exchange constant $ J_{\mathrm{RKKY}} = 9 \times 10^{-5} $ J/m$ ^2 $, which is in the typical range \cite{RKKY}. For the waveguide thickness we used
  $ t_{1,2} = 4 $ nm. A large enough magnetic field $ H_0 = 2 \times 10^5 $ A/m is applied along $ +y $ direction to bring the WGs to a remnant state.
 
 \section{Magnonic coupled wave-guide equations with spin-orbit torque}
  For  a deeper understanding of the full-fledge  numerical simulations presented below, it is instructive to formulate an analytical model by 
considering small deviations of $ \vec{m}_{s,p} = (\delta m_{x,p}, 0, \delta m_{z,p}) $ away from the  initial equilibrium $ \vec{m}_{0,p} = \vec{y} $.
Introducing  $ \psi_p = \delta m_{x,p} + i \delta m_{z,p} $ we deduce from linearizing Eq. (\ref{LLG})  the  coupled waveguide equations 
 \begin{equation}
 \begin{aligned}  
 i\frac{\partial \psi_1}{\partial t} - [(\omega_0 - \alpha \omega_J) - i(\omega_J + \alpha \omega_0)]\psi_1 + q \psi_2  &= 0, \\
 i\frac{\partial \psi_2}{\partial t} - [(\omega_0 + \alpha \omega_J) + i(\omega_J - \alpha \omega_0)]\psi_2 + q \psi_1  &= 0.
 \label{pt-equation}
 \end{aligned}              
 \end{equation}
 For convenience, we introduce in addition to the 
  coupling  strength  $ q = \frac{\gamma J_{\mathrm{RKKY}}}{(1+i \alpha)\mu_0 M_s t_p} $, the SOT coupling at zero intrinsic damping $\kappa= \gamma J_{\mathrm{RKKY}}/(2\mu_0 M_s t_p)=q|_{\alpha\to 0} $. The intrinsic frequency of the waveguides is given by 
  $ \omega_0 = \frac{\gamma}{1+\alpha^2}(H_0 + \frac{2 A_{\mathrm{ex}}}{\mu_0 M_s} k_x^2 + \frac{J_{\mathrm{RKKY}}}{2\mu_0 M_s t_p}) $ which is for the material studied here is in the GHz. Essential for the behavior akin to PT-symmetric systems  is the SOT-driven 
   gain-loss term $ \omega_J = \frac{\gamma c_J}{1+\alpha^2}  $.
   The wavevector along $x$ direction is  $ k_x $.
     Eq. (\ref{pt-equation}) admits a clear interpretation:  The magnonic  guided modes in  the first waveguide ( WG1)  are subject to the confining 
      complex potential  $V(z) = V_R(z) + i V_i (z) $ with $ V_R(z_0) = \omega_0 - \alpha \omega_J $ and $ V_i(z_0) = -\omega_J - \alpha \omega_0 $.
      In WG2 the potential is  $V_R(-z_0) = \omega_0 + \alpha \omega_J $ and $V_i(-z_0) = \omega_J - \alpha \omega_0 $.  The mode coupling is mediated by $q$ which determines the periodic magnon power exchange between  WG1 and WG2 in absence of SOT.
 
  For a PT symmetric system the condition  $ V_R(z_0) = V_R(-z_0) $ and $ V_i(z_0) = -V_i(-z_0) $ must apply, which is   obviously fulfilled if the intrinsic damping is very small  ($ \alpha \to 0 $).  Comparing the current and the photonic case, in the latter case the sign of the  imaginary part of the WGs refractive index  is tuned. Here we control with SOT the imaginary part of the permeability which we explicitly prove by deriving and analyzing  of the magnetic susceptibility (cf. Supp. Materials). This finding  points to a  new route for designing PT-symmetric magneto-photonic
  structures via permeability engineering.  We note, for a finite  magnetic damping $ \alpha $ a  PT-behavior is still viable as  confirmed by the full numerical simulations that we  discuss below. 
  \section{Magnon dynamics across the spontaneous PT-symmetry breaking transition}
The dispersion $ \omega(k_x) $ of the modes governed by Eq. (\ref{pt-equation}) reads 
   \begin{figure}
  	\includegraphics[width=0.5\textwidth]{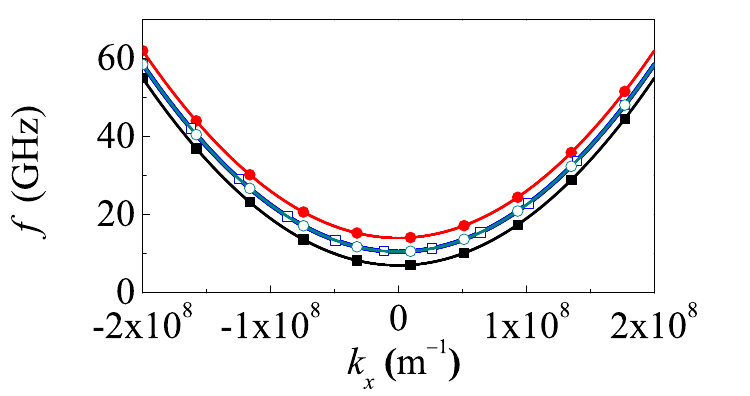}
  	\caption{\label{kx-w}  Merging of the acoustic  ($ \omega_J = 0 $, solid squares)  and optical magnon ($ \omega_J = 0 $, solid dots) modes dispersion
  		 $ \mathrm{Re}[\omega] (k_x) $ when approaching  the loss/gain-balanced exceptional point $\omega_J = \kappa $ (open dots).}
  \end{figure}
 %
 \begin{equation}
 \omega = (1-i \alpha)\omega_0 \pm \sqrt{q^2 - \omega_J^2 + 2i \alpha \omega_J^2 + \alpha^2 \omega_J^2}
 \label{eigenvalues}           
 \end{equation}
which describes both the acoustic and optical magnon modes \cite{PhysRevB.80.144425} and depends parametrically on $ \omega_J$ and $q$.
For $ \alpha\to 0 $ (in which case $q\equiv \kappa$) the eigenvalues are always real in the  PT-symmetric regime below the  gain/loss-balanced threshold $ \omega_J/\kappa < 1 $.
At the exceptional point  $ \omega_J/\kappa = 1 $, the two eigenvalues and eigenmodes become identical.
For  $ \omega_J/\kappa > 1 $ (by increasing the current density for instance) we enter the  PT-symmetry  broken phase, and the eigenvalues turn complex, as  typical for PT-symmetric systems \cite{PhysRevLett.101.080402,PhysRevLett.103.093902}.  The splitting between the two
 imaginary parts is determined by $2\kappa[(\omega_J/\kappa)^2 - 1]^{1/2}$ and is tunable by external fields. This fact is useful
 when exploiting the enhanced waveguides sensitivity to magnetic perturbations round the exceptional point.
  Allowing for a small damping  $ \alpha $  does not alter the modes behavior, as demonstrated by the full numerical results in a Fig. \ref{model}(b-c).  The full
  magnon dispersions ($ \mathrm{Re}[\omega] $ versus $ k_x $ curves ) for $ \omega_J/\kappa < 1 $ and $ \omega_J/\kappa = 1 $ are shown in Fig. \ref{kx-w}.
 The  symmetry of our waveguide brings in a special behavior of  the magnon signal transmission, meaning the propagation of 
 a superposition of  eigenmodes: Without charge current in the spacer ($\omega_J=0$), a signal injected at one end in one waveguide  oscillates  between WG1 and WG2 (due to the coupling $\kappa$) in a manner that is  well-established in coupled wave guide theory (cf.  Fig. \ref{model}(d) ).
 Switching on the charge current,  $\omega_J/\kappa$ becomes finite and the beating of the magnon power between WG1 and WG2 increases (cf.  Fig. \ref{model}(e)), as deducible from  Eqs. (\ref{pt-equation}), and also encountered  in optical wave guides  \cite{PhysRevLett.101.080402}.
Eqs. (\ref{pt-equation})  also indicate that near  the exceptional point,  a magnonic wavepacket injected in one waveguide
 no longer oscillates between the two waveguides but travels simultaneously  in both waveguides, as confirmed in
 Fig. \ref{model}(f) by full numerical simulations. This behavior resembles the optics case \cite{Ruter2010}.  We note that in our waveguides, this limit is simply achieved by tuning the external electric and magnetic fields that then change the ratio  $\omega_J/\kappa$. 
 We also found in line with Ref. [\citenum{Ruter2010}]  a non-reciprocal propagation  below the exceptional point.
  Passing  the exceptional point ($\omega_J/\kappa>1$) the magnonic signal  always propagates in the guide with gain and is quickly damped  in the guide with loss.   

\section{Enhanced sensing at PT-symmetry breaking transition}
 To assess the susceptibility of our setup to external magnetic perturbations we apply 
 an external microwave field $ \vec{h}_{p} $ which adds to effective field in   the LLG equation. In frequency space we deduced that 
 $ \widetilde{\psi}_p = \sum_{p\prime} \chi_{pp\prime} \gamma \widetilde{h}_{m,p\prime} $ 
 (tilde stands for Fourier transform),
 with $ h_{m,p} = h_{x,p} + i h_{z,p} $,  and $ \chi_{pp\prime} $  is the dynamic magnetic susceptibility which has the matrix  form 
 \begin{equation}
\begin{small}
\displaystyle \chi = \frac{1}{(\omega_{k}-i\alpha \omega - \omega) ^2+ \omega_c^2 - \kappa^2} \left( \begin{matrix} (\omega_k-i\alpha \omega) + (i \omega_c - \omega) & \kappa \\ \kappa & (\omega_k-i\alpha \omega) - (i \omega_c - \omega) \end{matrix} \right),
\label{chi-dynamic}
\end{small}
\end{equation}
with $ \omega_c = \gamma c_J $ and $ \omega_{k} = \gamma (H_0+ \frac{2 A_{\mathrm{ex}} k_x^2}{\mu_0 M_s} + \frac{J_{\mathrm{RKKY}}}{2\mu_0 M_s t_p})  $.

   \begin{figure}
	\includegraphics[width=0.85\textwidth]{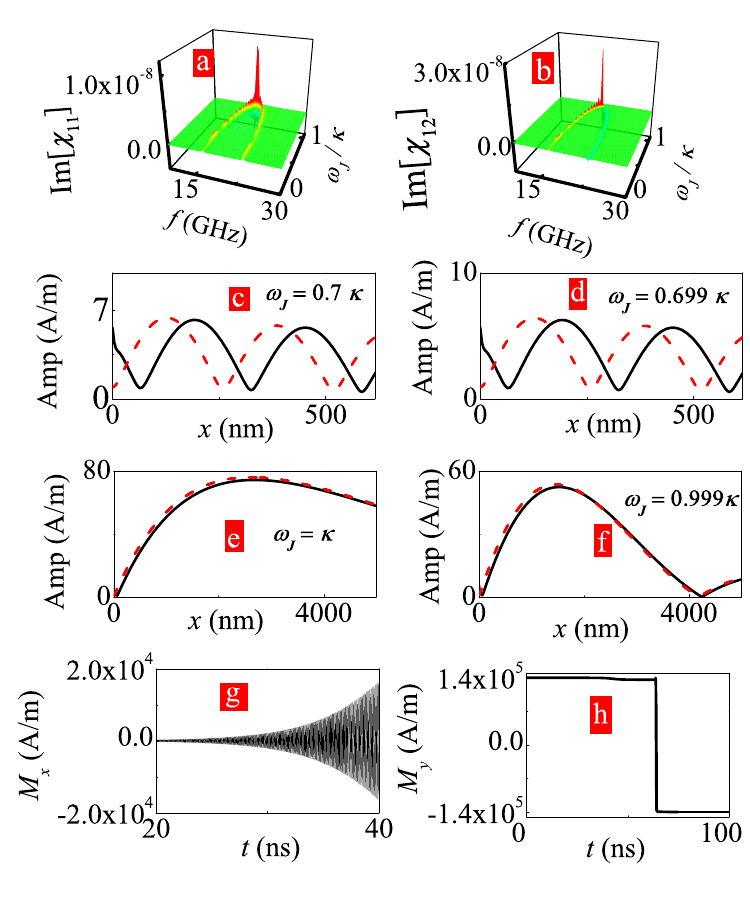}
	\caption{\label{chi-sense}  Magnetic susceptibility $ \mathrm{Im}[\chi_{11}] $ (a) and $ \mathrm{Im}[\chi_{12}] $ (b) as functions of $ f $ and $ \omega_J $. Peaks in $ \mathrm{Im}[\chi_{11}] $ and $ \mathrm{Im}[\chi_{12}] $ are found at the exceptional point $ \omega_J = \kappa $. (c-f) Exciting spin waves of frequency 20 GHz at $ x = 0$ in WG1, spatial profiles of spin wave amplitude for different $ \omega_J $. Black solid line and red dashed line represents the amplitudes in WG1 and WG2, respectively. Near $ \omega = \kappa $, a slight variation in $ \omega_J $ causes  marked  changes in the spin wave amplitudes (e-f), while the change is negligible near $ \omega = 0.7\kappa $ (c-d). (g-h) At $ \omega = \kappa $, increasing the external magnetic field ($ H_0 = 2 \times 10^5 $ A/m) in WG1 by 100 A/m, the time dependence of $ M_x $(g) and $ M_y $(h) at $ x = 2000 $ nm in WG2. }
\end{figure}

Near the exceptional point the system becomes strongly sensitive, for instance  to changes in the charge current term $ \omega_c $, 
as testified by the behavior of   the susceptibility which is  demonstrated  for the imaginary parts of  $ \chi_{11} $ and $ \chi_{12} $ in Fig. \ref{chi-sense}.
 The high sensitivity of the excited spin waves on $ \omega_c $ near the exception point (see Fig. \ref{chi-sense}(c-f)) is exploitable 
 to detect slight changes in charge current density $ c_J $. 

Furthermore, near the exceptional point, our setup is strongly sensitive to changes in the magnetic environment.  
As an example, at $ \omega_J = \kappa $, if the magnetic field $ H_0 $ (or local magnetization) is increased  by 100 A/m in WG1, large amplitude   spin-wave oscillations are generated in WG2, as evidenced by the time dependence of $ M_x(x = 2000 \mathrm{nm}) $  in WG2 (Fig. \ref{chi-sense}(g)). The spin wave amplification leads eventually to a reversal of  $ M_y $ in WG2  (Fig. \ref{chi-sense}(h)). Away from the PT-breaking transition, e.g. for  $ \omega_J = 0.7\kappa $,  when $ H_0 $ is reduced by the same amount 
 in WG1,  virtually no   changes  in propagating spin waves are observed (not shown).
 Obviously, this magnon amplification may serve as  a  tunable sensor for the magnetic environment.

\section{Current-induced switching in magnetic PT-symmetric   junctions}
 A special feature of magnetic systems is the possibility of current-induced switching (described by  Eq. (\ref{LLG}) but not by Eqs. (\ref{pt-equation})) 
 \cite{Baumgartner2017}.
   \begin{figure}
 	\includegraphics[width=0.85\textwidth]{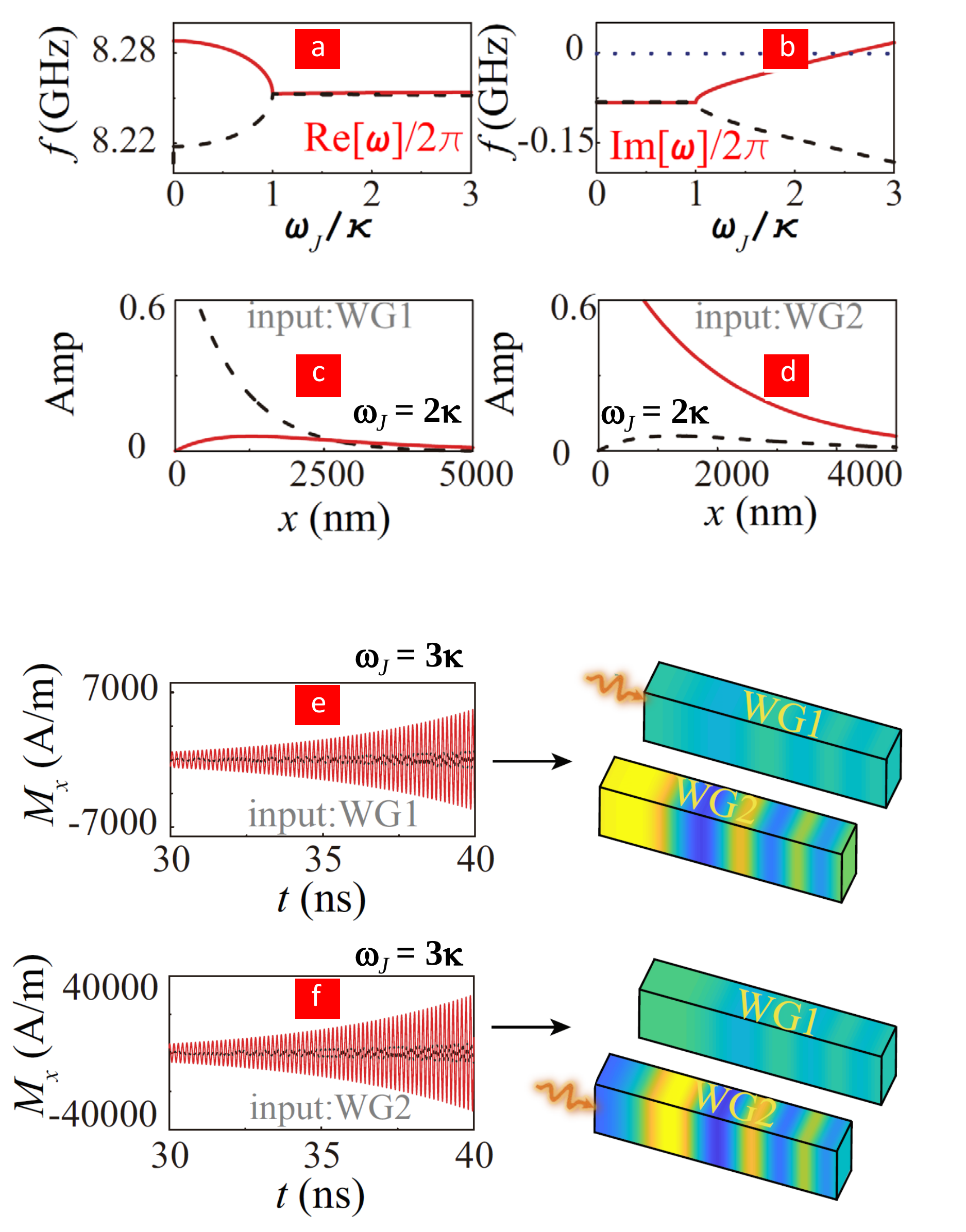}
 	\caption{\label{ce-w-smaller}  (a-b) Real and imaginary parts of the eigenmodes  $ \omega $ as varying the loss/gain balance by scanning $ \omega_J $ (meaning, the SOT strength). The wave vector is $ k_x = 0.03 $ nm$ ^{-1} $ and the intrinsic coupling between WG1 and WG2 $ \kappa $ is lowered, as compared to Fig. \ref{model}  (by choosing $ J_{\mathrm{RKKY}} = 9 \times 10^{-7} $ J/m$ ^2 $). (c-d)  Spatial profiles of magnon wave amplitudes (as normalized to their maxima)  for $ \omega_J = 2\kappa $,  and ($ \mathrm{Re}[\omega] = 2\pi \times 20 $ GHz). Black dashed lines corresponds to WG1 and red solid line to WG2. (e-f) Time dependence (at the location $ x = 2000 nm $) and the spatial profiles (at $ t = 40 $ ns) of the $ x $ component  of the magnetization $ M_x $
 		for $ \omega_J = 3\kappa $.
 		The color variation  from blue to red corresponds to  a  $ M_x $ change from the negative maximum to the positive maximum. }
 \end{figure}
In fact, for large current densities we are well above the exceptional point. In this case the magnetic system  becomes unstable towards switching. We find with further increasing the charge current density (enhancing  $ \omega_J $),   the local magnetization in  guide 2 is indeed  switched to  $ -y $. 
Magnon dynamics above the exceptional point is still possible  however by tuning the spacer material properties or its thickness to obtain a smaller $ \kappa $, for instance  with $ J_{\mathrm{RKKY}} = 9 \times 10^{-7} $ J/m$ ^2 $ and  $ \alpha = 0.01 $. In this case the condition $ \omega_J \gg \alpha \omega_0 $ is not satisfied anymore, and the influence of intrinsic magnetic losses ($ \alpha $) in both wave guides is important. Nonetheless, even without reaching the strict PT-symmetric condition, we still observe that the real parts of the two eigenvalues merge at the same point $ \omega_J = \kappa $, and the two imaginary parts become different when $ \omega_J > \kappa $, as shown by Fig. \ref{ce-w-smaller}(a-b). When $ \omega_J = 2\kappa $, the two imaginary parts are both negative, meaning that both modes are evanescent. The propagation of magnonic signal launched in one waveguide end is shown in Fig. \ref{ce-w-smaller} (c-d) evidencing that the spin waves in the two waveguides decay differently. An input signal   in the waveguide with enhanced damping leads to an  evanescent spin wave in WG1. Injecting the signal in  WG2, the attenuation of spin wave  is weaker, and its amplitude is always larger. When $ \omega_J = 3\kappa $ and $ \mathrm{Im}[\omega] $ of the optical magnon mode turns positive, we observe that  SOT induces spin wave amplification with time (Fig. \ref{ce-w-smaller}(e-f)). This finding  is  interesting   for  cavity optomagnonics \cite{Osada}.\\
For input signal in WG1 or WG2, the spin wave amplitude is always larger in WG2 with a negative effective damping. Also, the excited spin wave amplitude is much larger when the input is in the WG2. Thus, no matter from  which waveguide we start, the output signal is  always distributed at the end of WG2, a fact that can be employed for constructing magnonic logic gates.

    \begin{figure}
 	\includegraphics[width=0.85\textwidth]{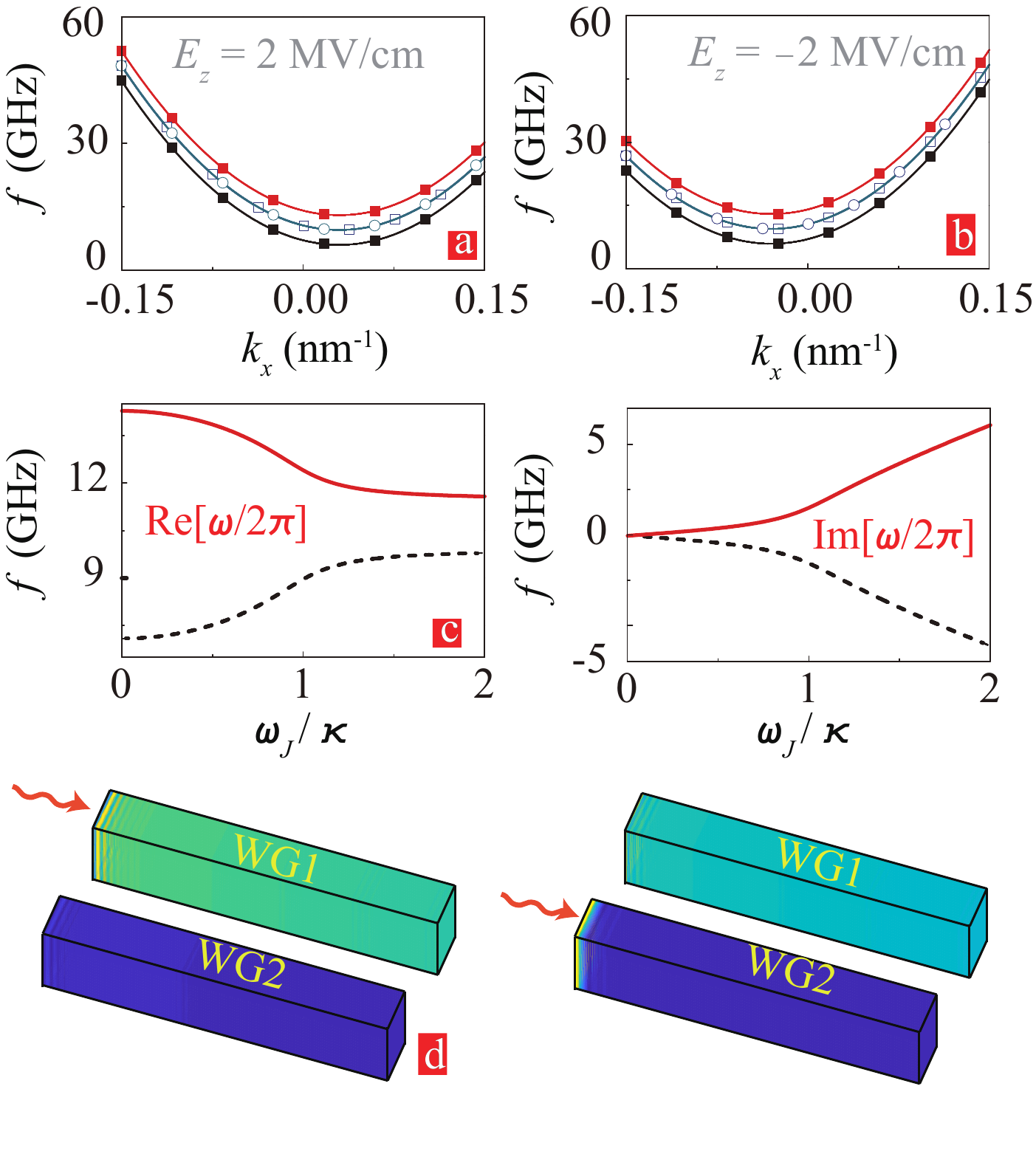}
 	\caption{\label{ce-w-dmi}
 		Control of coupled magnonic waveguide characteristics by an external electric field in presence of  Dzyaloshinskii-Moriya interaction. 
 		 (a-b) Applying a static electric field $ \vec{E}_{1,2}=(0,0,E_z) $ with $ E_z = \pm2 $ MV/cm to both waveguides modify the magnon dispersion $ \mathrm{Re}[\omega] ( k_x) $ curves for $ \omega_J = 0 $ (solid dots) and $\omega_J = \kappa $ (open dots). (c)  Real  and imaginary parts of two eigenmodes $ \omega $ as functions of $ \omega_J $ and in the presence of 
 		  two static  electric fields (or voltages) applied with opposite polarity to  the two waveguides 
 		  ($ \vec{E}_1=(0,0,E_z) $ and $ \vec{E}_2=(0,0,-E_z) $, and  $ E_z = 2 $ MV/cm) at $ k_x = 0.1 $ nm$ ^{-1} $. (d) 
 		  Spatial profiles of the propagating spinwave amplitudes when
 		  applying electric field in WG1 ($ \vec{E}_1=(0,0,E_z) $ with $ E_z = 2 $ MV/cm, and  $ \vec{E}_2=(0,0,0) $). Color scale from blue to red corresponds to  amplitude change from 0 to its maximum.}
 \end{figure}
\section{Dzyaloshinskii-Moriya interaction in electrically controlled PT-symmetric waveguides}
In  magnetic layers and at their interfaces  an antisymmetric exchange,  also called Dzyaloshinskii-Moriya (DM),  interaction
\cite{PhysRevLett.104.137203, PhysRevB.88.184404} may exist. 
In our context it is  particularly interesting that the DM interaction may allow for  a coupling 
to an external electric field  $ \vec{E} $ and voltage gates. 
  The contribution to the system free energy density in the presence of DM and 
   $ \vec{E} $ is $ E_{elec} = -\vec{E}\cdot \vec{P} $, with the spin-driven polarization   $ \vec{P} = c_E [(\vec{m} \cdot \nabla) \vec{m} - \vec{m}(\nabla \cdot \vec{m})] $ \cite{PhysRevApplied.12.034015, PhysRevLett.113.037202}. This   alters  the magnon dynamics through the additional term $ \vec{H}_{elec} = -\frac{1}{\mu_0 M_s} \frac{\delta E_{elec}}{\delta \vec{m}} $ in the effective field $ \vec{H}_{\mathrm{eff}} $.
   To uncover the role of DM interaction on the magnon dynamics in PT symmetric waveguides we consider three cases:
     (i) The two waveguides  experience the same static electric field $ \vec{E}_{1,2}=(0,0,E_z) $.\\
      (ii)  The electric fields in the two waveguides are opposite to each other, i.e. $ \vec{E}_1=(0,0,E_z) $ and $ \vec{E}_2=(0,0,-E_z) $. \\
      (iii) The electric field is  applied only to waveguide 1. These situations can be achieved by electric gating. \\
  For the case (i) with $ \vec{E}_{1,2}=(0,0,E_z) $, $ \omega_0 = \frac{\gamma}{1+\alpha^2}(H_0 - \frac{2 c_E E_z k_x}{\mu_0 M_s} + \frac{2 A_{\mathrm{ex}}}{\mu_0 M_s} k_x^2 + \frac{J_{\mathrm{RKKY}}}{2 \mu_0 M_s t_p}) $ in Eq. \ref{pt-equation}, and the condition for PT-symmetry still holds. Applying an electric field along the $ z $ axis causes an asymmetry in the magnon dispersion. As shown by Fig. \ref{ce-w-dmi}, the positive $ E_z $ shifts the dispersion towards positive $k_x$  while a negative $ E_z $ shifts it in the opposite direction. With increasing $ \omega_J $, the changes of Re[$ \omega $] and Im[$ \omega $] (not shown) are similar to these in Fig. \ref{model}(b-c).

 As for the case $ \vec{E}_1=(0,0,E_z) $ and $ \vec{E}_2=(0,0,-E_z) $,  in the two equations (\ref{pt-equation})  $ \omega_0 $ is different. Explicitly: $ \omega_0 = \frac{\gamma}{1+\alpha^2}(H_0 \mp \frac{2 c_E E_z k_x}{\mu_0 M_s} + \frac{2 A_{\mathrm{ex}}}{\mu_0 M_s} k_x^2 + \frac{J_{\mathrm{RKKY}}}{2\mu_0 M_s t_p}) $  where the  $ - $ sign applies for WG1 and the $ + $ sign corresponds to WG2. Hence, under an asymmetric electric field the
   potential  $ V_R $ is not even ($ V_R(z_0) \ne V_R(-z_0) $), and the PT-symmetry condition can not be satisfied. The $ \omega_J $ dependence of Re[$ \omega $] and Im[$ \omega $] are shown in Fig \ref{ce-w-dmi}, and no exceptional point can be strictly identified in this case.

   For case (iii), we set $ \vec{E}_1=(0,0,E_z) $ and  $ \vec{E}_2=(0,0,0) $.  The  PT-symmetry condition is not satisfied. When  the electric field is  applied only to a single guide, it  shifts selectively the magnon dispersion relation in this guide. Therefore, the magnon wave in the lower frequency range   propagates solely in the guide with the electric field. As shown in Fig. \ref{ce-w-dmi}(d), we excite the  magnonic wavepacket with a frequency in the WG1 or WG2, the magnonic wave always propagates in the waveguide 1 which amounts to a magnon channeled  by the electric field, while the propagation in the other guide is suppressed. This example illustrates  yet another handle to steer magnonic waves swiftly and at low energy consumption by pulsed electric gating.
  
  \section{Conclusions}
  Magnonic waveguides based on magnetic junctions that exhibit a transition from a PT-symmetric to a PT-symmetry broken phase  may act near the transition (exceptional) point as effective sensor for changes in external fields and in the magnetic environments and also serve  as magnonic amplifier or magnetic switch.  The particular behavior of the waveguides magnetic susceptibility is also reflected in the permeability (cf. supplementary materials) pointing to a new route  to  PT-symmetric  magneto-photonics. 
   Magnonic propagation is highly controllable by  external electric and magnetic fields that can derive the system across the exceptional point and lead to  controlled power distribution in the waveguides as well as non-reciprocal or amplified magnon waves. 
  DM interaction allows for dispersion engineering via external electric fields, and for PT-symmetry based large-amplitude spin excitations.
  These  observations underline the potential of PT-symmetric  magnonics as the basis for  additional   
  functionalities  of  magnetophotonic,  spintronics and cavity magnonic devices that are highly controllable by external parameters.

\begin{acknowledgement}

This research is  financially  supported by the DFG through SFB 762 and SFB TRR227, National Natural Science Foundation of China (No. 11704415, 11674400, 11374373), and the Natural Science Foundation of Hunan Province of China (No. 2018JJ3629).

\end{acknowledgement}


\textbf{Author contributions}\\
XGW performed all numerical simulations and analytical modeling. JB conceived and supervised the project.  XGW and JB wrote the paper .
XGW , JB, and GHG discussed, interpret and agreed on the content.\\

\textbf{Competing interests: }\\
The authors declare no competing financial and non-financial interests.\\
\textbf{Full data availability statements:}
All technical details for producing the figures are enclosed in the supplementary materials. Data are available from the authors upon request.

\begin{suppinfo}
	
	\begin{itemize}
		\item pt-symmetry-supp.pdf: Numerical simulation details, dynamic magnetic permeability in separated waveguides, and the influence of intrinsic damping.
	\end{itemize}
	
\end{suppinfo}

\bibliography{pt-nanolett}
\end{document}